\documentclass[conference,letterpaper]{IEEEtran}

\usepackage[T1]{fontenc}
\usepackage[cmex10]{amsmath} 
\usepackage[cmintegrals]{newtxmath}
\usepackage{bm} 

\usepackage{enumitem}
\usepackage{graphicx}
\usepackage{microtype}
\usepackage[space,noadjust]{cite}
\usepackage{subcaption}
\usepackage{soul,xcolor}
\usepackage[ruled,vlined]{algorithm2e}
\SetAlFnt{\small}
\SetAlCapFnt{\small}
\SetAlCapNameFnt{\small}

\usepackage{etoolbox}
\makeatletter
\patchcmd{\@begintheorem}{\textit}{\textbf}{}{}
\patchcmd{\@begintheorem}{\itshape}{\bfseries}{}{}
\makeatother

\newcommand{\absent}{\mathbb{U}^\text{abs}}

\newtheorem{theorem}{Theorem}
\newtheorem{corollary}{Corollary}
\newtheorem{lemma}{Lemma}

\newtheorem{definition}{Definition}

\newtheorem{remark}{Remark}
\newtheorem{example}{Example}

\graphicspath{{.}{../pictures/}{pictures/}{../../pictures/}}

\def\gap{1.05ex}
\abovedisplayskip\gap
\belowdisplayskip\gap
\abovedisplayshortskip\gap
\belowdisplayshortskip\gap

\begin{document}

\title{Optimal-Rate Characterisation for\\ Pliable Index Coding using Absent Receivers}

\author{\IEEEauthorblockN{Lawrence Ong}\thanks{This work is supported by the ARC Future Fellowship FT140100219 and by NSF grants CNS-1526547 and CCF-1815322.}
\IEEEauthorblockA{University of Newcastle\\
  Email: lawrence.ong@newcastle.edu.au}
\and

\IEEEauthorblockN{Badri N.\ Vellambi}
\IEEEauthorblockA{University of Cincinnati\\
Email: badri.vellambi@uc.edu}

\and

\IEEEauthorblockN{J\"{o}rg Kliewer} 
\IEEEauthorblockA{New Jersey Institute of Technology\\
Email: jkliewer@njit.edu}

}

\IEEEoverridecommandlockouts
\maketitle

\begin{abstract}
We characterise the optimal broadcast rate for a few classes of pliable-index-coding problems. This is achieved by devising new lower bounds that utilise the set of \textit{absent} receivers to construct decoding chains with \textit{skipped messages}. This work complements existing works by considering problems that are not complete-$S$, i.e.,  problems considered in this work do not require that all receivers with a certain side-information cardinality to be either present or absent from the problem. We show that for a certain class, the set of receivers is critical in the sense that adding any receiver strictly increases the broadcast rate.
\end{abstract}

\section{Introduction}

Index coding studies the optimal coding and rate requirements in a network with a single sender and multiple receivers connected by a noiseless broadcast link. In index coding, the sender is assumed to have $m$ messages and each receiver \emph{knows} a subset of the $m$ messages and \emph{wants} a specific subset of messages it does not know. Index coding~\cite{baryossefbirk11,neelytehranizhang13,ongholim16,arbabjolfaeikim18trends}, its secure variant~\cite{dauskachekchee12,ongvellambiyeohklieweryuan2016}, and its connection to network coding~\cite{rouayhebsprintsongeorghiades10,effrosrouayheblangberg15,ongkliwervellambiyeoh-it-2018} have received significant research interest.

Recently, a variant of index coding, known as \emph{pliable index coding}, was introduced~\cite{brahmafragouli15}. In this pliable variant, each receiver is posited not to want a specific subset, but instead to want \emph{any} subset of $t$ messages it does not already know. This variant is natural in applications where the receiver is flexible in which unknown message it wants to receive. One such example is when multiple receivers are each seeking a picture of an object on the internet, but each does not particularly care for a specific picture of the object.

Brahma and Fragouli~\cite{brahmafragouli15} focused solely on linear codes for pliable-index-coding problems and established that the problem is NP-hard. Further, they showed that if each receiver has at least $s_\text{min}$ and at most $s_\text{max}$ messages, then $\min \{s_\text{max} + t, m - s_\text{min}\}$ is an upper bound on the minimum number of transmissions (which is referred to as the minimum \textit{broadcast rate}) required for each receiver to obtain $t$ additional messages. And this bound is tight if the sender knows only the number of messages (as opposed to the exact message sets) that each receiver knows.  For a general setup, they 
approximated the order of dependence of the minimum broadcast rate in the limit as the number of messages and receivers grows.

Song and Fragouli~\cite{songfragouli18} also restricted their analysis to linear codes to show that if receivers having every possible strict subset of the message set are present, then the sender needs to send all $m$ messages. This result of all receivers being present was further strengthened by Liu and Tuninetti~\cite{liutuninetti17} to all (including non-linear) pliable index codes.

Liu and Tuninetti~\cite{liutuninetti17} defined a class of  complete-$S$ pliable-index-coding problems, where  $S\subseteq \{0,\ldots, m-1\}$ is a parameter.
Given $S\subseteq \{0,\ldots, m-1\}$, a complete-$S$ problem consists of \textit{all} $\binom{m}{i}$ receivers each having a different combination of $i$ messages, for every $i \in S$. Focusing on the case that each receiver requires only one message (that is,  $t=1$), they showed that the minimum broadcast rate for any linear or non-linear pliable index code for a complete-$S$ problem with $S=\{0,\ldots, m-1\}\setminus\{s_\text{min} ,\ldots, s_\text{max}\}$ is precisely $|S|=m-s_\text{max}+s_\text{min}+1$.

They later \cite{liutuninetti18} derived tight lower bounds based on decoding chains and maximum acyclic induced subgraphs to show that the minimum broadcast rate for any linear or non-linear pliable index code for a complete-$S$ problem with $S=\{s_\text{min} ,\ldots, s_\text{max}\}$ and $t=1$ equals $\min \{s_\text{max} + 1, m - s_\text{min}\}$.  

Existing results on exact minimum broadcast rates were established for certain complete-$S$ problems.
This paper considers the case $t=1$ and problems that are in general not complete-$S$. We identify a new technique based on \textit{absent receivers} to construct \emph{decoding chains with skipped messages} to derive lower bounds on the minimum broadcast rate for all pliable-index-coding problems that are applicable to both linear and non-linear codes. When combined with matching transmission codes (upper bounds), we establish precisely the minimum broadcast rate for several classes of pliable-index-coding problems.

We also introduce a notion of \emph{critical} set of receivers; such sets of receivers are maximal in the sense that the addition of any new receiver (that is absent) strictly increases the broadcast rate. In other words, each critical set of receivers is a maximal set of receivers supported by a fixed broadcast rate.

\subsection{Problem Formulation}

We use the following notation: $\mathbb{Z}^+$ denotes the set of natural numbers, $[a:b] := \{a, a+1, \dotsc, b\}$ for $a,b\in\mathbb{Z}^+$ such that $a < b$, and $X_S = (X_i: i \in S)$ for some ordered set $S$.

Consider a sender having $m \in \mathbb{Z}^+$ messages, denoted by $X_{[1 : m]} = (X_1, \dots, X_m)$. Each message $X_i \in \mathbb{F}_q$  is independently and uniformly distributed over a finite field of size~$q$. There are $n$ receivers having distinct subsets of messages, which we refer to as side information. Each receiver is labelled by its side information, i.e., the  receiver that has messages $X_{H}$, for some $H \subsetneq [1 : m]$, will be referred to as receiver $H$ or receiver with side information $H$. The aim of the pliable-index-coding problem is to devise an encoding scheme for the sender and a decoding scheme for each receiver satisfying pliable recovery of a message at each receiver. 

Without loss of generality, the side-information sets of the receivers are distinct; all receivers having the same side information can be satisfied if and only if (iff) any one of them can be satisfied. Also, no receiver has side information $H = [1:m]$ because this receiver cannot be satisfied. So, there can be at most $2^m-1$ receivers present in the problem. A pliable index coding problem is thus defined uniquely by $m$ and the set $\mathbb{U} \subseteq 2^{[1:m]} \setminus \{[1:m]\}$ of all receiver side information present in the problem. Lastly, 
any receiver that is not present, i.e., receiver~$H \in 2^{[1:m]} \setminus (\{[1:m]\} \cup \mathbb{U})$, is said to be \textit{absent}.

\begin{example}
Let $m= 3$, and  $\mathbb U=\{\emptyset, \{1\}, \{2\}, \{1,2\}, \{2,3\}]\}$. Then, the receivers $\{3\}$ and $\{1,3\}$ are absent.
\end{example}

Given a pliable-index-coding problem with $m$ messages and receivers $\mathbb U$, a pliable index code of length $\ell \in \mathbb{Z}^+$ consists of
\begin{itemize}
\item an encoding function of the sender, $\mathsf{E}: \mathbb{F}_q^m \rightarrow \mathbb{F}_q^\ell$; and 
\item for each receiver $H\in\mathbb{U}$, a decoding function $\mathsf{D}_H: \mathbb{F}_q^\ell \times \mathbb{F}_q^{|H|} \rightarrow \mathbb{F}_q$, such that $\mathsf{D}_H(\mathsf{E}(X_{[1:m]}),X_H) = X_i$, for some $i \in [1:m]\setminus H$.
\end{itemize}

The above formulation requires the decoding of only one message at each receiver, similar to that in Liu and Tuninetti~\cite{liutuninetti17,liutuninetti18}. Lastly, the aim is to find the optimal broadcast rate for a particular message size $q$, denoted by $\beta_q := \min_{\mathsf{E}, \{\mathsf{D}\}} \ell$ and the optimal broadcast rate over all $q$, denoted by $\beta := \inf_q \beta_q$.

\begin{remark}
  All results in this paper will be derived for $\beta_q$ for all $q \in \mathbb{Z}^+$. Consequently, the results are also valid for $\beta$.
\end{remark}

\section{New Lower Bounds}

\subsection{An optimal-rate expression}

We first express a lower bound on the optimal broadcast rate for pliable index coding in terms of an equivalence notion for index coding.
Define \textit{decoding choice} $D$ as follows:
\begin{equation}
  D: \mathbb{U} \rightarrow [1:m], \text{ such that } D(H) \in [1:m] \setminus H.
\end{equation}
Here, $D(H)$ is the message decoded by receiver $H$. 


Let $\mathcal{P}_{m,\mathbb{U}}$ denote a pliable-index-coding problem with $m$ messages and a set of receivers $\mathbb{U}$. For a fixed decoding choice~$D$ for $\mathcal{P}_{m,\mathbb{U}}$, denote the problem by $\mathcal{P}_{m,\mathbb{U},D}$. This means any code for $\mathcal{P}_{m,\mathbb{U},D}$ is a pliable index code for $\mathcal{P}_{m,\mathbb{U}}$ with the restriction that $\mathsf{D}_H(\mathsf{E}(X_{[1:m]}),X_H) = X_{D(H)}$ for all $H \in \mathbb{U}$, and vice versa.
With an abuse of notation, let the optimal broadcast rate for $\mathcal{P}_{m,\mathbb{U},D}$ be $\beta_q(\mathcal{P}_{m,\mathbb{U},D})$.
We can establish the following:
\begin{lemma}\label{lemma:equivalence}
  $\beta_q (\mathcal{P}_{m,\mathbb{U}}) = \min_D \beta_q(\mathcal{P}_{m,\mathbb{U},D}).$
\end{lemma}

\begin{IEEEproof}
 Clearly,
  $\beta_q (\mathcal{P}_{m,\mathbb{U}}) \leq \beta_q(\mathcal{P}_{m,\mathbb{U},D})$
for all $D$ because any code for $\mathcal{P}_{m,\mathbb{U},D}$ is a code for $\mathcal{P}_{m,\mathbb{U}}$. Since the inequality must be tight for at least one $D$, we have Lemma~\ref{lemma:equivalence}.
\end{IEEEproof}

$\mathcal{P}_{m,\mathbb{U},D}$ is in fact an index-coding problem~\cite{neelytehranizhang13,ongholim16,arbabjolfaeikim18trends}, 
with a message set~$X_{[1:m]}$ and a receiver set~$\mathbb{U}$, where each receiver~$H \in \mathbb{U}$ has $X_H$ and wants $X_{D(H)}$.

From Lemma 1, $\beta_q (\mathcal{P}_{m,\mathbb{U}})$ can be obtained by evaluating the optimal broadcast rates $\beta_q(\mathcal{P}_{m,\mathbb{U},D})$ of index-coding problems~$\mathcal{P}_{m,\mathbb{U},D}$ for all $D$. However, the optimal broadcast rate for index coding is not known in general, and  the search space over all possible $D$ grows exponentially with $m$.

\subsection{A lower bound based on acyclic subgraphs}

Nonetheless, we will utilise Lemma~\ref{lemma:equivalence} to formulate a lower bound for pliable index coding using results for index coding. More specifically,
\begin{equation}
  \beta_q (\mathcal{P}_{m,\mathbb{U}})  \geq \min_D \phi_q(\mathcal{P}_{m,\mathbb{U},D}),
\end{equation}
where $\phi_q(\mathcal{P}_{m,\mathbb{U},D})$ is any lower bound on $\beta_q(\mathcal{P}_{m,\mathbb{U},D})$.

We now state a lower bound for index coding~\cite{neelytehranizhang13}, expressed through a directed-bipartite-graph representation of an index-coding problem. Any index-coding problem can be specified by a bipartite graph with these two disjoint, independent sets: the message node set and the receiver node set. A directed edge from receiver node~$r$ to message node~$m$ exists iff receiver~$r$ has $X_m$ as side information; a directed edge from message node~$m$ to receiver node~$r$ exists iff receiver~$r$ wants $X_m$. 

Now, we perform one or more of following pruning operations, as many times as desired: \textsf{(a)}~remove a message node and all its incoming and outgoing edges; \textsf{(b)}~remove a receiver node and all its incoming and outgoing edges; \textsf{(c)}~remove a message-to-receiver edge. After a series of pruning operations, remove all message nodes with no outgoing edge. Let the resultant bipartite graph be $G'$, and the number of message nodes left by $m(G')$. If $G'$ is acyclic (in the directed sense), then we have the following lower bound, which generalises the maximum-acyclic-induced-subgraph (MAIS) lower bound~\cite{baryossefbirk11}.
\begin{lemma}{\cite[Lem.~1]{neelytehranizhang13}} \label{lemma:index-coding-lower-bound} Consider an index-coding problem~$\mathcal{I}$ and its bipartite-graph representation~$G$.
  After a series of pruning operations, if the resultant graph~$G'$ is acyclic, then
  $\beta_q(\mathcal{I}) \geq m(G')$.
\end{lemma}

A pliable-index-coding problem~$\mathcal{P}_{m,\mathbb{U}}$ with a decoding choice~$D$---that is, the index-coding problem $\mathcal{P}_{m,\mathbb{U},D}$---can be described by the following bipartite graph:
\textsf{(a)}~message nodes~$i \in [1:m]$;
\textsf{(b)}~receiver nodes~$H \in \mathbb{U}$;
\textsf{(c)}~each receiver node~$H$ has an outgoing edge to every message node~$i \in H$  and an incoming edge from node~$D(H)$.
Name this graph~$G_D$, and let $G_D'$ denote the resultant graph after a series of pruning operations on $G_D$.

Using Lemmas~\ref{lemma:equivalence} and \ref{lemma:index-coding-lower-bound},  we obtain the following lower bound for pliable index coding:
\begin{lemma}\label{lemma:acyclic} [Lower bound]
  Consider a pliable-index-coding problem $\mathcal{P}_{m,\mathbb{U}}$, and a set of bipartite graphs $\{G_D\}$ formed by all possible decoding choices $D$. Perform pruning operations on each $G_D$ to obtain an acyclic $G_D'$.  Then,
  \begin{equation}
    \beta_q (\mathcal{P}_{m,\mathbb{U}}) \geq \min_D m(G_D').
  \end{equation}
\end{lemma}

\subsection{Constructing acyclic subgraphs using decoding chains with skipped messages}

To use Lemma~\ref{lemma:acyclic}, one needs to consider all $D$, perform pruning operations on each $G_D$ to get an acyclic graph $G_D'$, and count the remaining number of message nodes $m(G_D')$.
We will instead use a decoding-chain argument to obtain the required $m(G_D')$. The concept of decoding chains was used to prove the MAIS lower bound~\cite{baryossefbirk11} and its extension~\cite[Lem.~1]{neelytehranizhang13} for index coding, and lower bounds for certain pliable-index-coding problems~\cite{liutuninetti17,liutuninetti18}.

In this paper, we propose the a new approach to construct decoding chains by introducing \textit{skipped messages}, which is implemented in the following randomised algorithm:


\vspace*{-1.95ex}
\begin{algorithm}[h]
\SetKwInOut{Input}{input}
\SetKwInOut{Output}{output}

\Input{$\mathcal{P}_{m,\mathbb{U},D}$} 
\Output{A \textit{decoding chain} $C$ (a totally ordered set with a total order $\leq_C$) and  a set of \textit{skipped messages} $S$}
$C \leftarrow \emptyset$; \texttt{\scriptsize\color{blue} (initialise $C$)}\\
$S \leftarrow \emptyset$; \texttt{\scriptsize\color{blue} (initialise $S$)}\\
\While{$C \neq [1:m]$}{
  \If(\texttt{\scriptsize\color{blue} (receiver $C$ is absent)}){$C \notin \mathbb{U}$}{
    Choose any $a \in [1:m] \setminus C$; $^\#$\\ \texttt{\scriptsize\color{blue} ($a$ is called a skipped message)}\\
    $C \leftarrow (C \cup \{a\}$, with $i \leq_C a,$ for all $i \in C$); \texttt{\scriptsize\color{blue} (expand $C$)}\\
    $S \leftarrow S \cup \{a\}$; \texttt{\scriptsize\color{blue} (expand $S$)}
  }
  \Else (\texttt{\scriptsize\color{blue} (receiver $C$ is present)})
  {
    $C \leftarrow (C \cup \{D(C)\}$, with $i \leq_C D(C),$ for all $i \in C$);\\
    \texttt{\scriptsize\color{blue} (add the message that receiver $C$ decodes)}
    }
  }
\caption{An algorithm to construct a decoding chain with skipped messages}
\label{algo:chain}
\end{algorithm}
\vspace*{-1.95ex}

We say that the algorithm ``\textit{skips}'' a message $a$, whenever we execute the step marked \# for that message $a$. We will see later that the number of skipped messages is an important parameter characterising lower bounds. We say that the algorithm ``\textit{hits}'' a receiver $H$ whenever $C$ is updated as $C \leftarrow H$. If receiver~$H$ is absent, we say that it hits an absent receiver~$H$. Note that receiver $[1:m]$ cannot exist, so when the algorithm ends, $[1:m]$ is not considered an absent receiver being hit.

\begin{remark} \label{remark:decoding-chain}
We highlight some properties of Algorithm~\ref{algo:chain}:
\begin{enumerate}
\item For a fixed $D$, the only uncertainty in constructing a chain is the choice of skipped messages. So, $(C,S)$ is completely determined by $D$ and the choice of skipped messages.
\item If an absent receiver~$H$ is hit, then subsequently a message~$a \notin H$ will be skipped, and vice versa.
  So, we skip a message iff we hit an absent receiver.
  \item The algorithm always commences by hitting receiver~$\emptyset$ first.
\end{enumerate}
\end{remark}

For a fixed $\mathcal{P}_{m,\mathbb{U},D}$, any choice of skipped messages results in a pair of $(C,S)$. Let $\mathbb{C}$ be the set of all $(C,S)$ pairs, obtained by varying different skipped messages.  We have the following:
\begin{lemma} \label{lemma:pruning-lower-bound}
   For each $(C,S) \in \mathbb{C}$ derived from a given $\mathcal{P}_{m,\mathbb{U},D}$ (or equivalently, $G_D$), there exists a series of pruning operations on $G_D$ yielding an acyclic $G_D'$ with $m(G_D') = |C \setminus S| = m - |S|$.
\end{lemma}

\begin{IEEEproof}
  Remove from $G_D$ all present receivers not being hit in the algorithm, and their connected edges.
Let the elements of $C$ in the order of construction of $C$ be $c_1,c_2,\ldots, \underline{c_i}, \ldots, c_{|C|}$, that is, $c_i \leq_C c_j$ iff $i \leq j$, where underlined elements are present in $S$ as well. By construction, if $c_i$ is underlined, then receiver~$\{c_1, \dotsc, c_{i-1}\}$ is absent. So, for each $c_i$ in $C$ that is not underlined, receiver~$\{c_1, \dotsc, c_{i-1}\}$ is present and has been hit in the algorithm, and therefore remains. This includes receiver~$\emptyset$ if $c_1$ is not underlined. So,  $|C \setminus S|$ receivers remain.
Next, remove all messages in $S$ (and their associated edges) so that only messages in $C \setminus S$ remain.

After these pruning operations, the graph $G_D'$ consists of the following edges for each remaining receiver node~$H$: \textsf{(a)}~outgoing edges from $H$ to all message nodes $i \in H \setminus S$, \textsf{(b)}~incoming edge from message node $D(H)$ to $H$. Also, by construction, for each remaining receiver node~$H$, $i \leq_C D(H)$ for all $i \in H$.

For $a \leq_C b$, we say that $b$ is \textit{larger} than $a$ in $C$, and $a$ is \textit{smaller} than $b$ in $C$.
  In $G_D'$, all edges flow from message nodes that are larger in $C$ to message nodes that are smaller in $C$, through receiver nodes. Hence, $G_D'$ is acyclic. Also, since each message node that remains is requested by a receiver that remains, no message node is removed after the pruning operations. So, $G_D'$ contains $|C \setminus S|$ message nodes. As $C$ contains all the messages $[1:m]$, we have  $|C \setminus S| = m - |S|$.
\end{IEEEproof}

\subsection{A lower bound via decoding chains with skipped messages}

We can express the lower bound in Lemma~\ref{lemma:pruning-lower-bound} as follows:

\begin{lemma} \label{lemma:chain-lower-bound} [Lower bound]
  Consider a pliable-index-coding problem $\mathcal{P}_{m,\mathbb{U}}$ and its bipartite-graph representation $G$.
  \begin{equation}
    \beta_q (\mathcal{P}_{m,\mathbb{U}}) \geq  m - \max_D  \min_{(C,S) \in \mathbb{C}} |S|. \label{eq:chain-lower-bound}
  \end{equation}
\end{lemma}

\begin{IEEEproof}
  From Lemmas~\ref{lemma:acyclic} and \ref{lemma:pruning-lower-bound}, we know that $\beta_q (\mathcal{P}_{m,\mathbb{U}}) \geq \min_D  (m - |S|)$, for any $(C,S) \in \mathbb{C}$ for each decoding choice $D$. By optimising $(C,S) \in \mathbb{C}$ for each $D$, we get Lemma~\ref{lemma:chain-lower-bound}.
\end{IEEEproof}

\begin{remark} \label{remark:max-min}
  Although the lower bound \eqref{eq:chain-lower-bound} involves minimising over all $(C,S) \in \mathbb{C}$, it is clear that any choice of $(C,S)$ for each $D$ will also give us a lower bound. Having said that, maximising over all $D$ is compulsory.
\end{remark}


\subsection{A lower bound based on nested chains of  absent receivers}
Denote the set of  absent receivers by $\absent := 2^{[1:m]} \setminus (\{[1:m]\} \cup \mathbb{U})$.

\begin{lemma} \label{lemma:necessary-nested}
  If an instance of Algorithm~\ref{algo:chain}  skips $L \in \mathbb{Z}^+$ messages,
  then there exists a \textit{nested chain} of absent receivers of length $L$, that is, $H_1 \subsetneq H_2 \subsetneq \cdots \subsetneq H_L$, with each $H_i \in \absent$. 
\end{lemma}

\begin{IEEEproof}
  A decoding chain $C$ is constructed by adding messages one by one. So, any receiver that is hit must contain all previously hit receivers. From Remark~\ref{remark:decoding-chain}, we know that if the algorithm skips $L$ messages, it must hit $L$ absent receivers, and these absent receivers must form a nested chain.
\end{IEEEproof}

We will now prove another lower bound that is easier to use compared to Lemma~\ref{lemma:chain-lower-bound} in some scenarios (for example, case~2 in Theorem~\ref{theorem:nesting} and Theorem~\ref{theorem:perfectly-nested}).

\begin{lemma} \label{lemma:simplier-lower-bound} [Lower bound]
  Consider a pliable-index-coding problem $\mathcal{P}_{m,\mathbb{U}}$ and its bipartite-graph representation $G$.
  Let $L \in \mathbb{Z}^+$ be the maximum length of any nested chain constructed from receivers absent in  $\mathcal{P}_{m,\mathbb{U}}$. We have that
 $\beta_q(\mathcal{P}_{m,\mathbb{U}}) \geq m-L$.
\end{lemma}

\begin{IEEEproof}
  $L$ must be the largest number of skipped messages evaluated over all decoding choices~$D$ and skipped-message sets. Otherwise, from Lemma~\ref{lemma:necessary-nested}, we have a nested chain of absent receivers of length $L+1$, which is a contradiction. Thus, 
$\displaystyle  m- L = m - \max_D  \max_{(C,S) \in \mathbb{C}} |S| \leq m - \max_D \min_{(C,S) \in \mathbb{C}} |S| \stackrel{\eqref{eq:chain-lower-bound}}{\leq} \beta_q(\mathcal{P}_{m,\mathbb{U}}).$
\end{IEEEproof}


\section{Criticality and Monotonicity}

Before we characterise the optimal broadcast rate of certain classes of pliable-index-coding problems, we introduce the notion of \textit{critical} receivers for pliable index coding.

In index coding, it is well-known that removing any message from the side information of any receiver cannot decrease the optimal broadcast rate $\beta$. Hence, the side-information sets of all receivers are said to be critical if removing any messages therein results in a strictly larger $\beta$.

However, in pliable index coding, removing messages from side-information sets may increase or decrease $\beta$. We will establish this in Corollary~\ref{corollary:criticality} later. Hence, criticality should not be defined for the messages in side-information sets. However, we can define criticality of pliable index coding with respect to the receivers. By noting that any pliable index code for $\mathcal{P}_{m,\mathbb{U}}$ is also a pliable index code for $\mathcal{P}_{m,\mathbb{U}^-}$, we have the following:
\begin{lemma} \label{lemma:monotonicity}
Let $\mathbb U^{-}\subseteq \mathbb U$. Then, $\beta_q (\mathcal{P}_{m,\mathbb{U}^-}) \leq \beta_q(\mathcal{P}_{m,\mathbb{U}})$.
 \end{lemma}


In light of this, we define the following.
\begin{definition}
  For pliable-index-coding $\mathcal{P}_{m,\mathbb{U}}$, the set of receivers $\mathbb{U}$ is said to be \textit{critical} iff adding any receiver to $\mathbb{U}$ strictly increases $\beta_q$. 
\end{definition}

So, for pliable index coding, critical receivers can be seen as a maximal receiver set that a broadcast rate can support. This is different from index coding, where critical side information can be seen as the minimal side information that is required to maintain a broadcast rate.

\section{Results on Optimal Broadcast Rates}
We now derive $\beta_q$ for a few classes of pliable-index-coding problems. 
For lower bounds, we use Lemma~\ref{lemma:chain-lower-bound} and Lemma~\ref{lemma:simplier-lower-bound} for different settings.
For achievability, we will engage \textit{cyclic codes} defined as follows. A cyclic code for messages $\{X_1, X_2, \dotsc, X_L\}$ is $(X_1 + X_2, X_2 + X_3, \dotsc, X_{L-1} + X_L) \in \mathbb{F}_q^{L-1}$. For notational convenience, we let the cyclic code for a single message $X_i$ be nil (that is, sending nothing).

\begin{theorem} \label{theorem:incomplete}
 Let $\mathcal{P}_{m,\mathbb{U}}$ be such that $|\absent| \neq 0$ and
  \begin{equation}
   \textstyle \mathop{\bigcup}\limits_{H \in \absent} H \neq [1:m]. \label{eq:union-not-full}
  \end{equation}
  Then $\beta_q(\mathcal{P}_{m,\mathbb{U}}) = m-1$.
\end{theorem}

\begin{IEEEproof}
  If receiver $\emptyset \in \mathbb{U}$, we remove it to get another pliable-index-coding problem $\mathcal{P}^- = \mathcal{P}_{m,\mathbb{U}\setminus \{\emptyset\}}$. Using Lemma~\ref{lemma:monotonicity}, $\beta_q(\mathcal{P}^-) \leq \beta_q(\mathcal{P}_{m,\mathbb{U}})$.

  We run Algorithm~\ref{algo:chain} on $\mathcal{P}^-$. Since receiver $\emptyset$ is missing, we start by skipping some message $a \in [1:m]$. We choose any $a \in [1:m] \setminus \mathop{\bigcup}_{H \in \absent} H$, which is possible due to \eqref{eq:union-not-full}. After this step, for \textit{any} decoding choice $D$, Algorithm~\ref{algo:chain} must terminate without skipping any more messages (meaning that it will not hit any absent receiver). This is because $a$ (which is included in $C$ in the first step) is not in the side-information set of any absent receiver. So, Algorithm~\ref{algo:chain} terminates with  $S = \{a\}$.

  Invoking Lemma~\ref{lemma:chain-lower-bound}, we have $\beta_q(\mathcal{P}^-) \geq m-1$. Note that we need not minimise the algorithm over all $(C,S)$ here; see Remark~\ref{remark:max-min}. This completes the lower bound.

  For achievability, pick any $H \in \absent$. We send $X_H$ uncoded, and $X_{[1:m]\setminus H}$ using a cyclic code. This gives a codelength of $m-1$. Note that any receiver that does not have all messages in $H$ as side information will be able to decode a new message. Also, any receiver that has all messages in $H$ must also have at least one (but not all) messages in $[1:m] \setminus H$---because receiver $H$ is absent---and hence it can decode a new message from the cyclic code.
\end{IEEEproof}

It has been shown~\cite{liutuninetti17} that if all receivers are present, then $\beta_q = m$. We now strengthen the result to if and only if.
\begin{theorem}
  $\beta_q(\mathcal{P}_{m,\mathbb{U}}) = m$ iff $\mathbb{U} = 2^{[1:m]}\setminus \{[1:m]\}$.
\end{theorem}

\begin{IEEEproof}
  We only need to prove the ``only if'' part. Equivalently, we show that if $\mathbb{U} \neq 2^{[1:m]}\setminus \{[1:m]\}$, then $\beta_q(\mathcal{P}_{m,\mathbb{U}}) \neq m$. We start by observing that if $\mathbb{U} \neq 2^{[1:m]}\setminus \{[1:m]\}$, then at least one receiver must be absent. By letting the absent receiver be $H$, we have $\mathbb{U} \subseteq 2^{[1:m]}\setminus \{ [1:m], H\} := \mathbb{U}^+$. As $H \neq [1:m]$, we have 
$\beta_q(\mathcal{P}_{m,\mathbb{U}}) \leq \beta_q (\mathcal{P}_{m,\mathbb{U}^+}) = m-1$, where the inequality follows from Lemma~\ref{lemma:monotonicity} and the equality from Theorem~\ref{theorem:incomplete}
\end{IEEEproof}

We now present our results to absent receivers $\absent$ in some cases where $\mathop{\bigcup}_{H \in \absent} H = [1:m]$.

\begin{theorem} \label{theorem:nesting}
  Consider a pliable-index-coding problem $\mathcal{P}_{m,\mathbb{U}}$. If any of the following is true, then $\beta_q(\mathcal{P}_{m,\mathbb{U}}) = m-1$.
  \begin{enumerate}
  \item (no nested absent pair) $J \nsubseteq K$, for all distinct $J, K \in \absent$.
  \item (one nested absent pair) $J \subsetneq K$, for exactly one pair of $J, K \in \absent$.
  \end{enumerate}
\end{theorem}

\begin{IEEEproof}
Theorem~\ref{theorem:incomplete} covers the case $\mathop{\bigcup}_{H \in \absent} H \neq [1:m]$. So, in the proof, we consider only $\mathop{\bigcup}_{H \in \absent} H = [1:m]$.
  
  For achievability, we use the coding scheme for Theorem~\ref{theorem:incomplete}, that is, we choose any $H \in \absent$, and then send $X_H$ uncoded, and $X_{[1:m]\setminus H}$ using a cyclic code. This gives a code of length $m-1$. Note that this code works for the case where only receiver~$H$ is missing, will therefore works for the case where $H$ and more receivers are missing.

  For lower bounds, we start with case~1. Since no pair of absent receivers are nested, using Lemma~\ref{lemma:simplier-lower-bound}, we obtain the required lower bound $m-1$. 

For case~2, as there is a pair of nested absent receivers, Lemma~\ref{lemma:simplier-lower-bound} gives a loose lower bound of $m-2$. Suppose that receiver~$\emptyset$ is absent, then $J=\emptyset$, and only one another receiver $K$ can be absent, since the presence of any other absent receiver will yield at least two pairs of nested absent receivers. In this setting then, $\mathop{\bigcup}_{H \in \absent} H = \emptyset \cup K \neq [1:m]$, and by Theorem~\ref{theorem:incomplete}, we see that $\beta_q(\mathcal{P}_{m,\mathbb{U}}) = m-1$. 

Now, suppose that case 2 holds and $\emptyset$ is present. With $\emptyset \in \mathbb{U}$, we know that Algorithm~\ref{algo:chain} can start without skipping the first message to be included in $C$. We split the decoding choices into three sub-cases, and skip specific messages to avoid $|S|=2$.\\
  \underline{Sub-case 1:} $D$ such that the decoding chain does not hit any absent receiver. For this case, $|S|=0$.\\
  \underline{Sub-case 2:} $D$ such that the decoding chain first hits any absent receiver $H \neq J$. Then, we arbitrarily skip one message, and will not hit another absent receiver, since every receiver that has $H$ as a subset is present.  This gives $|S|=1$.\\
  \underline{Sub-case 3:} $D$ such that the decoding chain first hits $J$. Then, we skip a message $a \in [1:m] \setminus K$. We will not hit another absent receiver, as every receiver that has $J \cup \{a\}$ as a subset is present. This results in $|S|=1$. \\
  Maximising $|S|$ over all $D$, we get the lower bound $m-1$
\end{IEEEproof}

For the next result, we need first define a class of pliable-index-coding problems.
\begin{definition}
  A pliable-index-coding problem is said to have \textit{perfectly $L$-nested absent receivers} iff the messages $[1:m]$ can be partitioned into $L+1 \in [2:m]$ subsets $P_0, P_1, \dotsc, P_{L}$ (that is, $\mathop{\bigcup}_{i=0}^L P_i = [1:m]$ and $P_i \cap P_j = \emptyset$ for all $i \neq j$), such that only $P_0$ can be an empty set, and there are exactly $2^L-1$ \textit{absent} receivers, which are
  \begin{equation}
 \textstyle    P_0 \cup \left( \mathop{\bigcup}\limits_{i \in Q} P_i \right), \text{ for each } Q \subsetneq [1:L].
  \end{equation}
\end{definition}

\begin{figure}[t]
  \centering
  \includegraphics[scale=0.39]{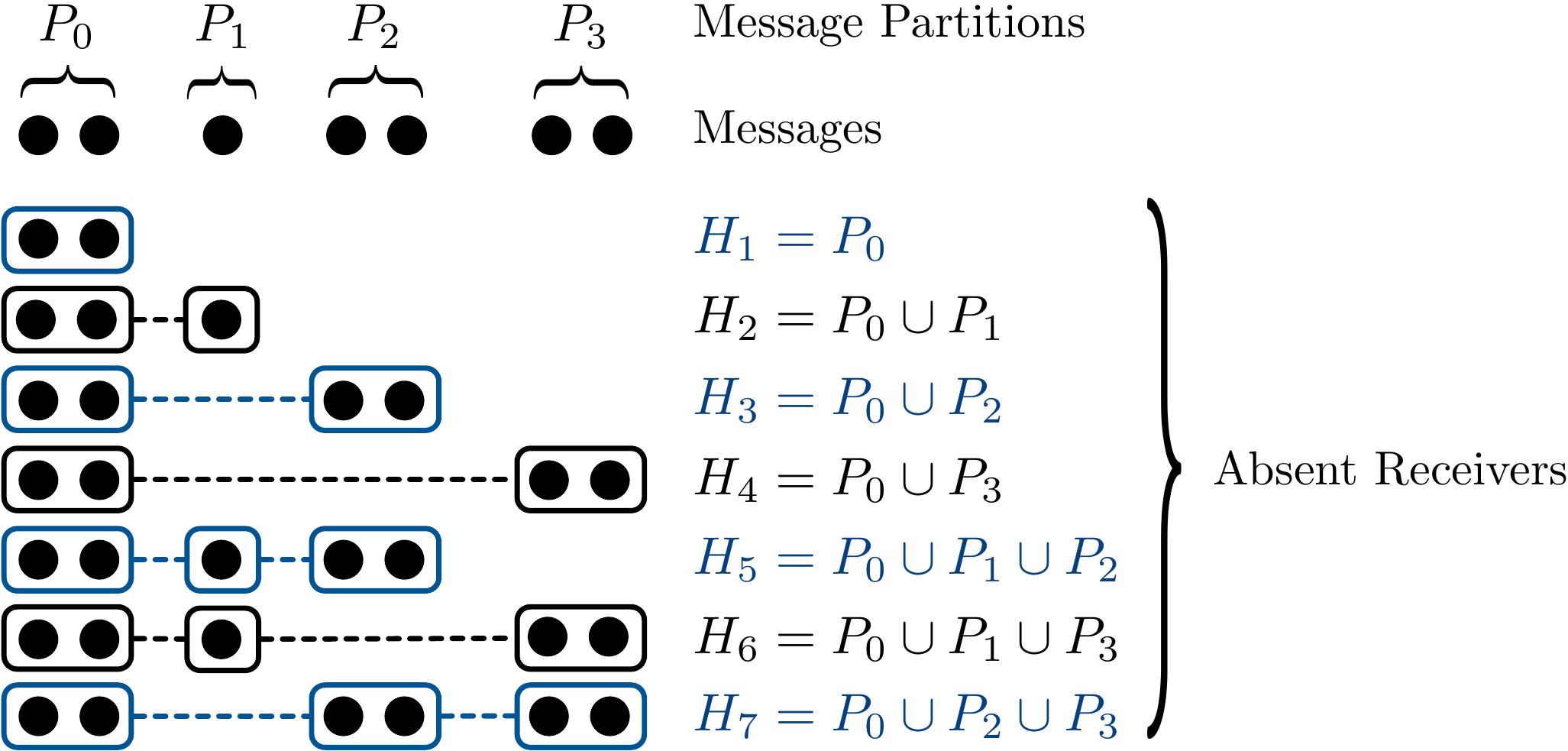}
  \caption{Perfectly 3-nested absent receivers, where circles denote messages, $\{P_i\}_{i=0}^3$ partitions, and $\{H_i\}_{i=1}^7$ absent receivers.}
  \label{fig:nested}
  \vspace{-0.94em}
\end{figure}

Figure~\ref{fig:nested} depicts an example of perfectly 3-nested absent receivers.

\begin{theorem} \label{theorem:perfectly-nested}
  For any pliable-index-coding problem $\mathcal{P}_{m,\mathbb{U}}$ with perfectly $L$-nested absent receivers, $\beta_q(\mathcal{P}_{m,\mathbb{U}}) = m-L$.
\end{theorem}

\begin{IEEEproof}
  For achievability we send $X_{P_0}$ uncoded and $X_{P_i}$ for each $i \in [1:L]$ using a cyclic code. One can verify that decodability of each present receiver can be satisfied.

  Since the maximum length of any nested chain of absent receivers is $L$, Lemma~\ref{lemma:simplier-lower-bound} gives  $\beta_q(\mathcal{P}_{m,\mathbb{U}}) \geq m-L$.
\end{IEEEproof}

\begin{lemma}
If $\absent$ is set of perfectly $L$-nested absent receivers, then $\mathbb{U}$ is critical.
\end{lemma}

\begin{IEEEproof}
  Start with $\mathcal{P}_{m,\mathbb{U}}$ with perfectly $L$-nested absent receivers.
  Imposed by the structure of $\mathbb{U}$, the maximum length of any nested chain of absent receivers is $L$, and they each must be in the following form:
  \vspace*{-1.7ex}
  \begin{multline}
    \textstyle  P_0 \subsetneq\,  P_0  \hspace{-0.25mm}\cup\hspace{-0.25mm} P_{i_1} \,\subsetneq\, P_0 \hspace{-0.25mm}\cup\hspace{-0.25mm} P_{i_1}  \hspace{-0.25mm}\cup\hspace{-0.25mm} P_{i_2}\subsetneq \dotsm \subsetneq\, P_0\hspace{-0.25mm}\cup\hspace{-0.25mm} \left(\mathop{\bigcup}\limits_{j=1}^{L-1} P_{i_j}\right), \label{eq:L-chain}
  \end{multline}
  for some distinct $i_1, \dotsc, i_{L-1}\in [1:L]$.

  We need to show that if we augment $\mathbb{U}$ with any absent receiver $H = P_0 \cup ( \mathop{\bigcup}_{i \in Q} P_i )$ for some $Q \subsetneq [1:L]$, then $\beta_q(\mathcal{P}_{m,\mathbb{U}^+}) \geq m-L+1$, where $\mathbb{U}^+ = \mathbb{U} \cup \{H\}$ is the  set of receivers after augmenting $H$.
  
  Clearly, if $H = P_0$, then receiver $P_0$ is no longer absent, and \eqref{eq:L-chain} is not possible. So, the  maximum length of any nested chain constructed from receivers absent in $\mathcal{P}_{m,\mathbb{U}^+}$ is $L-1$. Using Lemma~\ref{lemma:simplier-lower-bound}, we have $\beta_q(\mathcal{P}_{m,\mathbb{U}^+}) \geq m-L+1$.

  Otherwise, without loss of generality, let $H = P_0 \cup ( \mathop{\bigcup}_{1 \leq j \leq Q} P_j )$ for $Q \in [1:L-1]$. We will use Lemma~\ref{lemma:chain-lower-bound} to show that for any decoding choice $D$, we can construct a series of skipped messages $S$ such that $|S| \leq L-1$.

  Since in any attempt to skip $L$ messages, chain \eqref{eq:L-chain} is necessary, we only need to consider all decoding choices for which the first absent receiver being hit is $P_0$. After this, we choose to skip any message in $P_1$. Again, in the attempt to skip $L$ messages, the decoding choice must be made such that the next absent receiver being hit is $P_0 \cup P_1$. Repeating this, in iteration~$i \in [1:Q-1]$, after hitting each $P_0 \cup (\mathop{\bigcup}_{1 \leq j \leq i} P_j )$, we choose to skip any message in $P_{i+1}$. The next absent receiver being hit must then be $P_0 \cup (\mathop{\bigcup}_{1 \leq j \leq i+1} P_j )$, except when we reach $i+1 = Q$, where the absent receiver $P_0 \cup ( \mathop{\bigcup}_{1 \leq j \leq Q} P_j )$ has been included in $\mathbb{U}^+$. In this case, either \textsf{(a)} the next absent receiver being hit is $P_0 \cup ( \mathop{\bigcup}_{1 \leq j \leq Q} P_j ) \cup P_k$ for some $k \in [Q+1:L]$ if $Q \leq L-2$, or \textsf{(b)} the decoding chain terminates without hitting another absent receiver if $Q=L-1$. In any case, by the choice of skipped messages we devised, the maximum skipped messages is $L-1$ for any decoding choice, and therefore Lemma~\ref{lemma:chain-lower-bound} gives $\beta_q(\mathcal{P}_{m,\mathbb{U}^+}) \geq m-L+1$.
\end{IEEEproof}

With the above results, we can show the following:
\begin{corollary}\label{corollary:criticality} 
For any $\mathcal{P}_{m,\mathbb{U}}$, removing a message from a present receiver $H \in \mathbb{U}$ may strictly increase or strictly decrease the optimal broadcast rate $\beta_q(\mathcal{P}_{m,\mathbb{U}})$.
\end{corollary}

We prove Corollary~\ref{corollary:criticality} using the example below:

\begin{example} \label{example:increase-decrease}
  Consider $m=5$ and a set of absent receivers~$\absent_1 = \{ \{1,2,3\}, \{3\}, \{3,4\}\}$. Using Theorem~\ref{theorem:incomplete}, we have $\beta_q(\mathcal{P}_{m,\mathbb{U}_1}) = m-1$. Now, we remove message~5 from a present receiver~$\{3,4,5\} \in \mathbb{U}_1$. This is equivalent to replacing the present receiver~$\{3,4,5\}$ with a new present receiver~$\{3,4\}$. We get $\absent_2 = \{ \{1,2,3\}, \{3\}, \{3,4,5\}\}$, which forms perfectly 2-nested absent receivers. Using Theorem~\ref{theorem:perfectly-nested}, $\beta_q(\mathcal{P}_{m,\mathbb{U}_2}) = m-2$. We continue by removing messages 2 and 4 from the present receiver $\{2,3,4\} \in \mathbb{U}_2$.  This replaces the present receiver $\{2,3,4\}$ with a new present receiver $\{3\}$. We get $\absent_3 = \{ \{1,2,3\}, \{2,3,4\}, \{3,4,5\}\}$. Using Theorem~\ref{theorem:nesting}, $\beta_q(\mathcal{P}_{m,\mathbb{U}_3}) = m-1$.
\end{example}





\end{document}